\documentclass[onecolumn,groupedaddress,showkeys]{article}
\usepackage{float}
\usepackage{graphicx}
\usepackage{amssymb}
\usepackage{authblk}
\usepackage{amsmath}
\usepackage{caption}
\usepackage{subcaption}
\usepackage{xcolor}
\usepackage[super]{cite} 
\usepackage[breaklinks]{hyperref}
\hypersetup{colorlinks,urlcolor=black,citecolor=black,linkcolor=black,filecolor=black}
\usepackage{breakurl}
\usepackage{anysize}
\usepackage[left=1cm,right=1cm,top=1cm,bottom=1cm,includehead,includefoot]{geometry}
\begin{document}

	\title{Effect of air pollution on the growth of diabetic population}
	\author{Sourav Chowdhury\thanks{email: chowdhury95sourav@gmail.com}\qquad Suparna Roychowdhury\thanks{email: suparna@sxccal.edu}\qquad Indranath Chaudhuri\thanks{email: indranath@sxccal.edu}}
	\affil{Department of Physics, St. Xavier's College (Autonomous)\\
		30 Mother Teresa Sarani, Kolkata-700016, West Bengal, India}

	\maketitle
	
\begin{abstract}		
	Diabetes mellitus is a disease which is currently a huge health hazard globally. The cases of diabetes had increased by a significant amount in past decades. Also it has been predicted that it will further increase in future. Diabetes depends on various factors like obesity, physical inactivity. Also diabetes can depend on various environmental issues. In this article, our main focus is to study the dependence of the diabetic cases on the air pollution. We have used the data for diabetic population and PM2.5 concentration in the air for five countries from 2010 to 2021. Here we have studied the correlation between the diabetic cases data and PM2.5 concentration data. Also, we have done the linear regression analysis to find whether  this correlation is statistically significant.
\end{abstract}
	
\section{Introduction}
	Diabetes mellitus is a disease which is affecting global health widely. According to IDF (International diabetes federation), the number of people with diabetes has increased from 151 million to 537 million from 2000 to 2021 worldwide \cite{idf_atlas}. Also, they have predicted that this number will shoot up to 783 million by 2045. A person can suffer from diabetes with various reasons like lesser insulin secretion from Pancreas (Type-1 diabetes) or insulin resistance in the body cells (Type-2 diabetes). Also there is another type of diabetes which occurs at the time of pregnancy is called gestational diabetes. Diabetes disrupts homeostasis of the glucose level in the blood and thus blood glucose level remains higher than the normal level. This causes various short term and long term problems in the body.
	
	There are various factors which are behind the diabetes. Physical inactivity, unhealthy life style, obesity, smoking are the leading cause of the diabetes \cite{BOLES20171026,campagna2019smoking,burgio2015obesity}. Genetic factors are also a reason of the diabetes \cite{burgio2015obesity}. However, various environmental effects also influences the growth of the diabetic cases. There are some studies which shows that the global temperature increase and various social factors are behind the increment of diabetic cases \cite{Blauwe000317,dong2019effect}. Air pollution is a leading factor which is affecting the environment and thus the health of world population. Due to rapid industrialization and globalization, the air pollution increased by significant amount in past decades. Only recently the governments around the world are taking steps to control the air pollution. However, it is still a major problem that the world is facing. The contradiction between going green and development of industries which helps to advancement of economy is ever constant and growing. There are various studies which shows that the air pollution has a major role behind type-2 diabetes mellitus and gestational diabetes \cite{review1,review2}. A polluted air contains PM2.5 particles which causes oxidative stress and as a result various problems like hypertension, asthma and insulin resistivity occurs \cite{BOWE2018e301,YU2020105880,lim2019air}. So it is important to study the effect of air pollution on the diabetic cases. Thus in this regard we propose to explore the dependence of ever growing diabetic cases with air pollution as a world health hazard.
	
	Our paper is arranged as follows: In section~\ref{data} the data and data source is discussed. Section~\ref{method} describes the methods of our work and section~\ref{result} shows the results of this work. Finally, the concluding remarks are given in section~\ref{conclusion}. 
 \section{Data}\label{data}
 	In this work, we have used 2010 to 2021 diabetes data from IDF (International diabetes federation) of various countries \cite{atlas_data}. The PM2.5 concentration data (in $\mu g/m^{3}$) is taken from World bank data center (2010-2017) \cite{world_bank} and IQAir website (2018-2021) \cite{IQAir}. An initial analysis has been done for five countries: India, China, France, Germany, and UK whose data is robust.

\section{Method and analysis}\label{method}
	Here, we have assumed that the increment of diabetic cases is directly correlated with the PM2.5 concentrations in the air. So, from IDF data we have calculated the yearly increment of the diabetic cases. Then we have created a scatter plot with the yearly increment of diabetic cases against PM2.5 concentration for various countries. The value of the correlation coefficient have been evaluated for each of these countries.
	
\section{Results}\label{result}
\begin{figure}[H]
		\centering
		\begin{tabular}{cc}
			\includegraphics[scale=0.62]{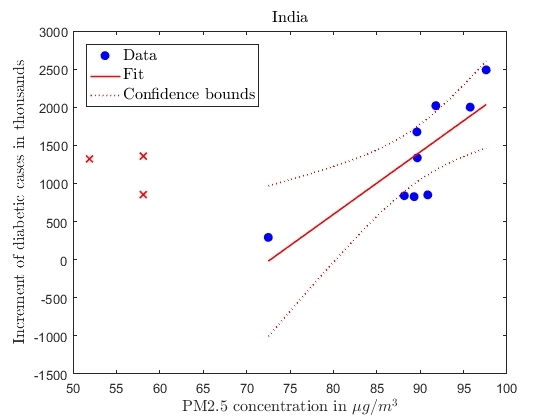}&
			\includegraphics[scale=0.62]{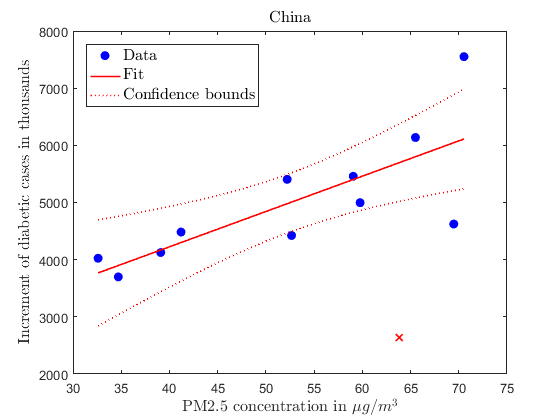}\\
			\includegraphics[scale=0.62]{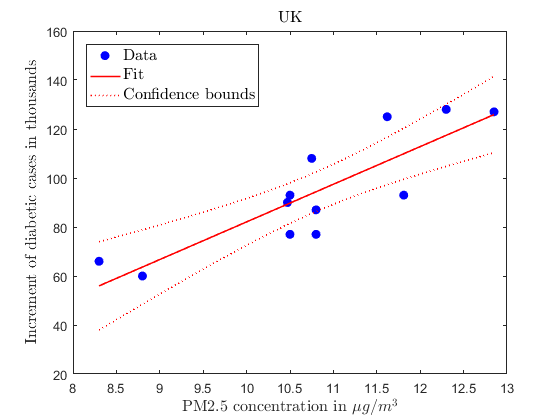}&
			\includegraphics[scale=0.62]{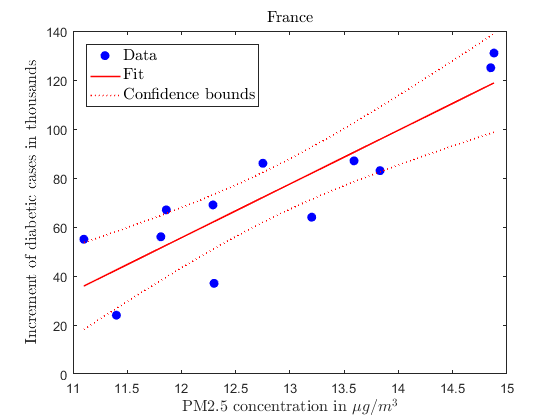}
		\end{tabular}
		\centerline{\includegraphics[scale=0.62]{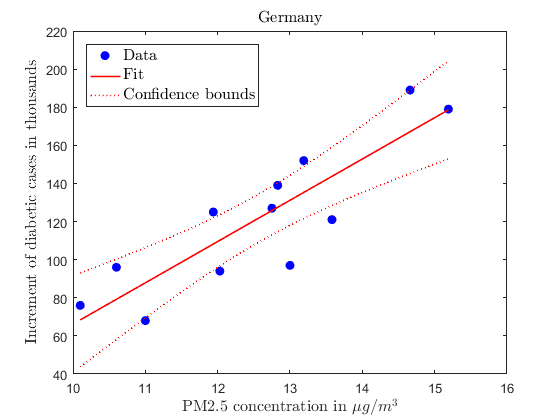}}
		\caption{{Regression model fit to the increment of diabetic cases with PM2.5 concentration data for various countries. Here red cross $(\times)$ denotes the outliers of the data.}\label{fig1}}
	\end{figure}

	In this section, we have shown various results of our work. The estimated values of the correlation coefficients $(r)$ of PM2.5 data and increment of the diabetic cases for different countries are shown in Table \ref{r_table}.
	\begin{table}[H]
		\centering
		\begin{tabular}{|c|c|}
			\hline
			Countries & Values of $r$ \\
			\hline
			India & 0.8037 \\
			China & 0.7668 \\
			UK & 0.8584 \\
			France & 0.8783 \\
			Germany & 0.8719\\
			\hline
		\end{tabular}
		\caption{{Values of correlation coefficients for different countries.}\label{r_table}}
	\end{table}
	It is generally known that if $r>0.8$ then the two variables are highly correlated and the correlation is positive \cite{stat}. From Table~\ref{r_table} it is seen that the $r$ values for most of the countries are greater than $0.8$. For China, $r$ value is slightly less than $0.8$. The linear regression model fit on the data are shown in Figure \ref{fig1}.
	
	In figure~\ref{fig1} the fitted regression model is plotted against the data with the confidence bounds. The outliers of the data are represented by red crossed points.	From this figure it is seen that the linear regression model fits well to the data. Different p-values of the linear regression fit for these countries are shown in Table~\ref{p_table}. 
	\begin{table}[H]
		\centering
		\begin{tabular}{|c|c|}
			\hline
			Countries & $p$-values \\
			\hline
			India & 0.00905 \\
			China & 0.0059 \\
			UK & 0.000351 \\
			France & 0.000171 \\
			Germany & 0.000218\\
			\hline
		\end{tabular}
		\caption{{p-values of correlation coefficients for different countries.}\label{p_table}}
	\end{table}
	If p-value $\leq 0.05$ then the correlation is statistically significant. Table~\ref{p_table} shows that p-values for all countries are less than 0.05. Thus the correlation between PM2.5 concentration and increment of diabetic cases is statistically significant.
	
\section{Conclusions}\label{conclusion}
	In this section, we have summarized the important features and results of our work. Here our main motive is to find the correlation between diabetes and air pollution. Thus we have taken data of diabetic cases and PM2.5 concentration from 2010 to 2021 for India, China, UK, France, Germany. We have calculated the value of correlation coefficient ($r$) between yearly increment of diabetic cases and PM2.5 concentration. We have found that the value of $r>0.8$ for most of the countries, which implies that these two variables are strongly correlated. Also we have applied the regression analysis on this data for different countries. For all of the countries, the $p<0.05$, which means that the correlation between increment of diabetic cases and PM2.5 concentration in the air is statistically significant. Thus we can say that air pollution affects our health and it is a significant reason behind the increment of diabetic cases around the world. There are various other factors which influences the increase of diabetic cases. In future, we aim to consider other factors like economic condition, obesity prevalence to understand their effects on the growth of diabetic cases globally. We also would like to include other countries in our analysis and increase the time span of our study to get a better idea of the environmental effects on diabetic cases.
	
\section*{Acknowledgment}

	The authors would like to thank the Department of Physics, St. Xavier's College, Kolkata for providing support during this work. One of the authors (S. C.) acknowledges the financial support provided from the University Grant Commission (UGC) of the Government of India, in the form of CSIR-UGC NET-JRF. 

\bibliographystyle{unsrt}
\bibliography{References}	

\begin{thebibliography}{10}

\bibitem{idf_atlas}
{\em International Diabetes Federation. IDF Diabetes Atlas, 10th edn. Brussels,
  Belgium: 2021.}
\newblock \url{https://www.diabetesatlas.org/}.

\bibitem{BOLES20171026}
Annette Boles, Ramesh Kandimalla, and P.~Hemachandra Reddy.
\newblock Dynamics of diabetes and obesity: Epidemiological perspective.
\newblock {\em Biochimica et Biophysica Acta (BBA) - Molecular Basis of
  Disease}, 1863(5):1026--1036, 2017.
\newblock Oxidative Stress and Mitochondrial Quality in Diabetes/Obesity and
  Critical Illness Spectrum of Diseases.

\bibitem{campagna2019smoking}
D~Campagna, A~Alamo, A~Di~Pino, C~Russo, AE~Calogero, F~Purrello, and R~Polosa.
\newblock Smoking and diabetes: dangerous liaisons and confusing relationships.
\newblock {\em Diabetology \& metabolic syndrome}, 11(1):1--12, 2019.
\newblock DOI: 10.1186/s13098-019-0482-2.

\bibitem{burgio2015obesity}
Ernesto Burgio, Angela Lopomo, and Lucia Migliore.
\newblock Obesity and diabetes: from genetics to epigenetics.
\newblock {\em Molecular biology reports}, 42(4):799--818, 2015.
\newblock DOI: 10.1007/s11033-014-3751-z.

\bibitem{Blauwe000317}
Lisanne~L Blauw, N~Ahmad Aziz, Martijn~R Tannemaat, C~Alexander Blauw, Anton~J
  de~Craen, Hanno Pijl, and Patrick C~N Rensen.
\newblock Diabetes incidence and glucose intolerance prevalence increase with
  higher outdoor temperature.
\newblock {\em BMJ Open Diabetes Research and Care}, 5(1), 2017.
\newblock DOI: 10.1136/bmjdrc-2016-000317.

\bibitem{dong2019effect}
Guangtong Dong, Lianlian Qu, Xuefeng Gong, Bing Pang, Weitian Yan, and Junping
  Wei.
\newblock Effect of social factors and the natural environment on the etiology
  and pathogenesis of diabetes mellitus.
\newblock {\em International Journal of Endocrinology}, 2019, 2019.

\bibitem{review1}
SA~Meo, AN~Memon, SA~Sheikh, FA~Rouq, A~Mahmood Usmani, A~Hassan, and SA~Arian.
\newblock Effect of environmental air pollution on type 2 diabetes mellitus.
\newblock {\em Eur Rev Med Pharmacol Sci}, 19(1):123--128, 2015.

\bibitem{review2}
Yongze Li, Lu~Xu, Zhongyan Shan, Weiping Teng, and Cheng Han.
\newblock Association between air pollution and type 2 diabetes: an updated
  review of the literature.
\newblock {\em Therapeutic Advances in Endocrinology and Metabolism},
  10:2042018819897046, 2019.
\newblock DOI: 10.1177/2042018819897046.

\bibitem{BOWE2018e301}
Benjamin Bowe, Yan Xie, Tingting Li, Yan Yan, Hong Xian, and Ziyad Al-Aly.
\newblock The 2016 global and national burden of diabetes mellitus attributable
  to pm2·5 air pollution.
\newblock {\em The Lancet Planetary Health}, 2(7):e301--e312, 2018.
\newblock DOI: 10.1016/S2542-5196(18)30140-2.

\bibitem{YU2020105880}
Guoqi Yu, Junjie Ao, Jing Cai, Zhongcheng Luo, Randall Martin, Aaron van
  Donkelaar, Haidong Kan, and Jun Zhang.
\newblock Fine particular matter and its constituents in air pollution and
  gestational diabetes mellitus.
\newblock {\em Environment International}, 142:105880, 2020.
\newblock DOI: 10.1016/j.envint.2020.105880.

\bibitem{lim2019air}
Chris~C Lim and George~D Thurston.
\newblock Air pollution, oxidative stress, and diabetes: a life course
  epidemiologic perspective.
\newblock {\em Current diabetes reports}, 19(8):1--14, 2019.
\newblock DOI: 10.1007/s11892-019-1181-y.

\bibitem{atlas_data}
{\em Diabetes data portal- IDF Diabetes atlas}.
\newblock \url{https://diabetesatlas.org/data/en/}.

\bibitem{world_bank}
{\em Air pollution information (from 2010 to 2017)- World bank data}.
\newblock \url{https://data.worldbank.org/indicator/EN.ATM.PM25.MC.M3}.

\bibitem{IQAir}
{\em Air pollution information (from 2018 to 2021)- IQAir}.
\newblock \url{https://www.iqair.com/world-most-polluted-countries}.

\bibitem{stat}
Roger~Purves David~Freedman, Robert~Pisani.
\newblock {\em Statistics}.
\newblock W. W. Norton \& Company, 2007.

\end{thebibliography}

\end{document}